\documentclass[prb,twocolumn,showpacs,preprintnumbers,amsmath,amssymb]{revtex4}
\usepackage{graphicx}
\usepackage{dcolumn}
\usepackage{bm}
\usepackage{graphicx}
\usepackage{epsfig}
\usepackage{epstopdf}
\include{graphic}

\begin{document}
\preprint{APS/123-QED}
\title{On the line shape of the electrically detected ferromagnetic resonance}

\author{M. Harder$^{1}$, Z. X. Cao$^{1,2}$, Y. S. Gui$^{1}$, X. L. Fan$^{1,3}$,
and C.-M. Hu$^{1}$\footnote{Electronic address: hu@physics.umanitoba.ca;
URL: http://www.physics.umanitoba.ca/$\sim$hu}}

\affiliation{$^{1}$Department of Physics and Astronomy, University
of Manitoba, Winnipeg, Canada R3T 2N2}

\affiliation{$^{2}$National Lab for Infrared Physics, Shanghai
Institute of Technical Physics, Chinese Academy of Science,
Shanghai 200083, People's Republic of China}

\affiliation{$^{3}$The Key Lab for Magnetism and Magnetic
Materials of Ministry of Education, Lanzhou University, Lanzhou
730000, People's Republic of China}

\date{\today}

\begin{abstract}

This work reviews and examines two particular issues related with
the new technique of electrical detection of ferromagnetic
resonance (FMR). This powerful technique has been broadly applied
for studying magnetization and spin dynamics over the past few
years. The first issue is the relation and distinction between
different mechanisms that give rise to a photovoltage via FMR in
composite magnetic structures, and the second is the proper
analysis of the FMR line shape, which remains the "Achilles heel"
in interpreting experimental results, especially for either
studying the spin pumping effect or quantifying the spin Hall angles via the
electrically detected FMR.

\end{abstract}

\keywords{Michelson interferometry, spintronics, electromagnetic
phase, spin resonances phase}

\pacs{85.75.-d, 75.40.Gb, 76.50.+g, 42.65.-k}


\maketitle

\section{Introduction}

Electrical detection of ferromagnetic resonance (FMR) in ferromagnets (FM) is a powerful new experimental tool which has transformed the research on spin and magnetization dynamics. \cite{Tsoi Nature2000, Ralph Nature2003, Tulapurkar Spindiode, VanWees spin-pumping, Saitoh ISHE, Kubota Spindiode, Sankey Spindiode, Gui SRE, Ralph ST, Gui PR, Azevedo spin-pumping, NanoFMR, VanWees PR, Yamaguchi SRE, Goennenwein PR, Nikolai2007PRB, Xiong Fe-FMR, Andre GaMnAs, Atsarkin LaSrMnO, Mosendz ISHE, Mosendz ISHE2,Liu SHE, Azevedo ISHE, Py/GaAs, Saitoh Y3Fe5O12/Pt ISHE, Hillebrands Y3Fe5O12/Pt ISHE, Gui Boundary, Gui Damping1, Gui Damping2, ND_Boone, DWR_Bedau, parametric excitation 1, parametric excitation 2} Over the past few years, this technique has generated a great deal of interest in the communities of magnetism, spintronics, and microwave technologies. It has been broadly applied for studying diverse material structures, ranging from ferromagnetic thin films such as Py (permalloy, Ni$_{80}$Fe$_{20}$),\cite{Gui PR, VanWees PR, Gui SRE, Yamaguchi SRE}, CrO$_2$,\cite{Goennenwein PR} Fe$_3$O$_4$,\cite{Goennenwein PR} single crystal Fe,\cite{Xiong Fe-FMR} GaMnAs,\cite{Andre GaMnAs} and La$_{1-x}$Sr$_x$MnO$_3$,\cite{Atsarkin LaSrMnO} bilayer devices such as Py/Pt,\cite{VanWees spin-pumping, Saitoh ISHE, Mosendz ISHE, Mosendz ISHE2, Liu SHE, Azevedo ISHE} Py/Au,\cite{Mosendz ISHE, Mosendz ISHE2} Py/GaAs,\cite{Py/GaAs} and Y$_3$Fe$_5$O$_{12}$/Pt,\cite{Saitoh Y3Fe5O12/Pt ISHE, Hillebrands Y3Fe5O12/Pt ISHE} to a variety of magnetic tunneling junctions (MTJ) based on magnetic multilayers.\cite{Tulapurkar Spindiode, Kubota Spindiode, Sankey Spindiode, NanoFMR} From a technical standpoint, its high sensitivity has made it possible to quantitatively determine spin boundary conditions\cite{Gui Boundary} and to directly measure non-linear magnetization damping\cite{Gui Damping1, Gui Damping2, ND_Boone}, the quasiparticle mass for the domain wall\cite{DWR_Bedau}, the phase diagram of the the spin-transfer driven dynamics\cite{Ralph Nature2003} and various kinds of parametric spin wave excitation \cite{Ralph Nature2003, parametric excitation 1, parametric excitation 2}. Its capability to probe the interplay of spins, charges, and photons has been utilized for studying spin rectification\cite{Gui SRE, Nikolai2007PRB}, spin pumping\cite{VanWees spin-pumping}, spin torque\cite{Ralph ST}, and spin Hall effects\cite{Mosendz ISHE,Liu SHE,Azevedo ISHE}, which have led to the proposing and realization of novel dynamic spintronic devices such as the spin battery,\cite{RMP2005, Wang spin-pumping, VanWees spin-pumping} spin diode,\cite{Tulapurkar Spindiode, Kubota Spindiode, Sankey Spindiode} spin dynamo,\cite{Gui SRE, Nikolai2007PRB} and spin demodulator \cite{Demodulation}. Very recently, its ability to detect coherent processes\cite{Phase Andre, Phase Fan, Phase Zhu} has enabled electrical probing of the spin-resonance phase and the relative phase of electromagnetic waves\cite{Phase Andre}, which pave new ways for microwave sensing\cite{Bai2008}, non-destructive imaging,\cite{Phase Andre} and dielectric spectroscopy\cite{Phase Zhu}. Such a coherent capability is especially exciting as it resembles the latest achievement in semiconductor spintronics, where a new platform for coherent optical control of spin/charge currents has been developed by using nonresonant quantum interferences. \cite{Zhao Coherent, Wang Coherent, Werake Coherent}

From the physical standpoint, many different effects may generate a time-independent dc voltage in magnetic materials via the FMR. Reported mechanisms involve spin rectification\cite{Gui SRE, Nikolai2007PRB}, spin pumping\cite{VanWees spin-pumping}, spin torque\cite{Ralph ST}, spin diode\cite{Tulapurkar Spindiode, Kubota Spindiode, Sankey Spindiode}, spin Hall\cite{Liu SHE} and inverse spin Hall effects\cite{Saitoh ISHE, Mosendz ISHE, Mosendz ISHE2, Azevedo ISHE}. Two major issues stand out here: (1) A unified picture clarifying the relations and distinctions between such diverse mechanisms has not been established, which leads to increasing controversy and confusion in interpreting and understanding experimental results. A stunning example of this issue is found in the very recent studies of the spin Hall effect via electrically detected FMR, where two similar experiments performed on similar devices were interpreted completely differently.\cite{Mosendz ISHE, Liu SHE} (2) When more than one mechanism simultaneously plays a role in the FMR generated dc voltage, proper interpretation requires a quantitative analysis of the FMR line shape. In our opinion, this has remained the "Achilles heel" in recent studies of spin pumping and the spin Hall effect which utilize electrically detected FMR. The purpose of this article is to address these two critical issues with a brief review of the key physics of this subject, followed by systematically measured experimental data with detailed theoretical analysis.

This paper is split into three main sections. First we provide a brief review of different mechanisms which may generate the photovoltage via the FMR. Then we use the dynamic susceptibility obtained from a solution of the Landau-Lifshitz-Gilbert equation to derive analytical formulae for analyzing the line shape and the symmetry properties of the photovoltage generated through spin rectification. Finally we present experimental results measured from different samples, at different frequencies, and in different experimental configurations, showing that the FMR line shape is determined by the relative phase of microwaves which is sample and frequency dependent.

\section{A Brief Review of Electrical Detection of FMR}

\begin{figure} [t]
\begin{center}
\epsfig{file=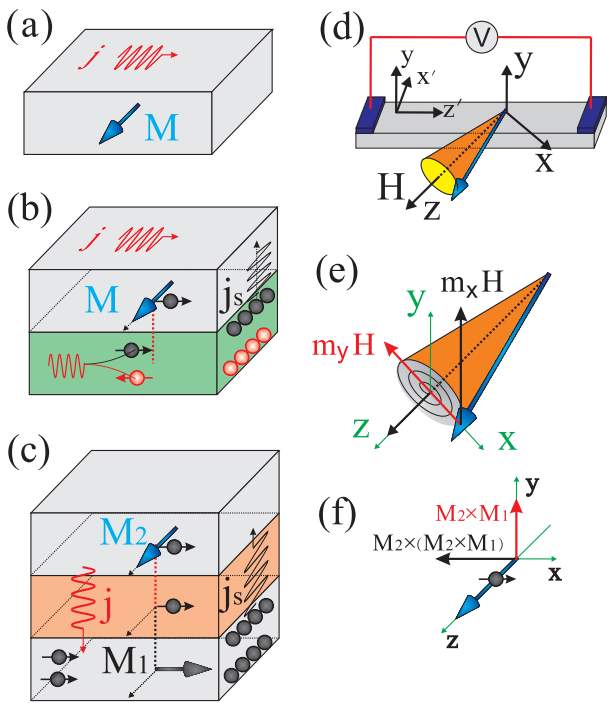,width=8 cm} \caption{(color online). Dynamic response of magnetic structures under microwave irradiation: (a) Single thin film layer where the spin rectification is due to the magnetic field torque as shown in (e). (b) Magnetic bilayer device which has two rf currents $\textbf{j}$ and $\textbf{j}_s$ with different spin polarizations. Therefore spin rectification is due to both magnetic field and spin torques. (c) Magnetic tunneling junction with both $\textbf{j}$ and $\textbf{j}_s$.  (d) Coordinate system for single ferromagnetic microstrips measured in this work under an in-plane applied static magnetic field $\textbf{H}$.  The $z^\prime$-axis is fixed along the strip and the direction of current flow, while the $z$-axis is rotated to follow the direction of $\textbf{H}$. (e) Components of magnetic field torque.  (f) Spin torque in magnetic tunneling junction.}  \label{fig1}
\end{center}
\end{figure}

Under microwave excitation at angular frequency $\omega$, the rf electric ($\textbf{e}$) and magnetic ($\textbf{h}$) fields inside a ferromagnetic material can be described as $\textbf{e}=\textbf{e}_{0}e^{-i\omega t}$ and $\textbf{h}=\textbf{h}_{0}e^{-i(\omega t-\Phi)}$, respectively. Note that in general, due to the inevitable losses of microwaves propagating inside the ferromagnetic material, there is a phase difference $\Phi$ between the dynamic $\textbf{e}$ and $\textbf{h}$ fields. Such a relative phase is determined by the frequency-dependent wave impedance of the materials\cite{Jackson}. As shown in Fig. \ref{fig1}, the rf $\textbf{e}$ field drives a rf current $\textbf{j}=\sigma \textbf{e}$, while the rf $\textbf{h}$ field exerts a field torque on the magnetization and drives it to precess around its equilibrium direction [Fig. \ref{fig1}(e)]. Such a magnetization precession is described by the non-equilibrium magnetization $\textbf{m}=\hat{\chi}\textbf{h}$. Here $\sigma$ and $\hat{\chi}$ are the high-frequency conductivity and Polder tensor, respectively. Note that due to the resonance nature of the precession, $\textbf{m}$ lags $\textbf{h}$ by a spin resonance phase $\Theta$. However, despite the phase of $\Phi$ and $\Theta$, the dynamic $\textbf{j}$ and $\textbf{m}$ keep the coherence of their respective driving fields, so that the product of any combination of their components may generate a time independent signal proportional to $\langle Re(\tilde{j})\cdot Re(\tilde{m})\rangle$, where $\langle\rangle$ denotes the time average. The amplitude of such a signal depends on the phase difference of $\textbf{j}$ and $\textbf{m}$, which can be easily understood from the trigonometric relation: $\cos(\omega t) \cdot \cos(\omega t - \Phi)=[\cos(\Phi)+\cos(2\omega t-\Phi)]/2$. This is the spin rectification\cite{Gui SRE} as we highlight in Table I. For transport measurements on magnetic structures under microwave irradiation, various magnetoresistance effects such as anisotropic magnetoresistance (AMR), giant magnetoresistance (GMR) and tunneling magnetoresistance (TMR) make corrections to Ohm's law via their corresponding magnetoresistance terms\cite{Juretschke, Nikolai2007PRB}. Such non-linear terms typically lead to the product of $\textbf{j}$ and $\textbf{m}$. Spin rectifications induced by such magnetoresistance effects are listed in Table I by the terms labeled $V_{MR}$. The general feature of $V_{MR}$ is that its amplitude depends on both the relative phase $\Phi$ and the spin resonance phase $\Theta$, which leads to a characteristic phase signature of the FMR line shape\cite{Phase Andre, Phase Zhu}.

Similar to the effect of the rf $\textbf{h}$ field torque, a spin torque induced by a spin polarized current may also drive magnetization precession. For example, in a bilayer [Fig. \ref{fig1}(b)] made of a ferromagnetic layer and a nonmagnetic layer with spin-orbit coupling\cite{Liu SHE}, in addition to the rf current $\textbf{j}$ flowing in the ferromagnetic layer, the rf $\textbf{e}$ field also induces a rf charge current flowing in the nonmagnetic layer. Via the spin Hall effect in such a nonmagnetic layer with spin-orbit coupling, the rf charge current can be converted into a spin current $\textbf{j}_{s}$, which may flow into the ferromagnetic layer and then drive the magnetization precession via the spin torque. Such a spin torque induced non-equilibrium magnetization can be described by $\textbf{m}=\hat{\chi}_{j}\textbf{j}_{s}$, where the spin-torque susceptibility tensor $\hat{\chi}_{j}$ introduces a spin resonance phase $\vartheta$ that is different from $\Theta$ in $\hat{\chi}$. Following a similar consideration for the magnetoresistance induced spin rectification, a photovoltage depending on the spin Hall effect may be generated in the ferromagnetic layer. This is the physical origin of the spin Hall induced spin rectification effect,\cite{Liu SHE} which is listed in Table I by the term labeled $V_{SH}$. In MTJ [Fig. \ref{fig1}(c)], the spin polarized current $\textbf{j}_{s}$ can be directly generated in the ferromagnetic layer where the magnetization is pinned along a different direction than that of the free layer. It tunnels into the free layer and drives the magnetization precession via the spin torque [Fig. \ref{fig1}(f)]. The induced spin rectification signal has been measured in spin diodes\cite{Tulapurkar Spindiode, Kubota Spindiode, Sankey Spindiode}, which is listed in Table I by the term labeled $V_{SD}$.

Over the past few years, systematic studies on spin rectifications induced by the field ($V_{MR}$) and spin torque ($V_{SH}$, $V_{SD}$) have been performed, respectively, at the University of Manitoba\cite{Gui SRE,Nikolai2007PRB,Xiong Fe-FMR,Andre GaMnAs,Gui Boundary,Gui Damping1,Gui Damping2,Phase Andre,Phase Zhu,Bai2008} and Cornell University\cite{Ralph Nature2003,NanoFMR,Sankey Spindiode,Ralph ST,Liu SHE,Kupferschmidt2006}. It has been found that due to the coherent nature of spin rectification, $V_{MR}$, $V_{SH}$ and $V_{SD}$ all depend on the phase difference between $\textbf{j}$ and $\textbf{m}$. However, only the field torque spin rectification ($V_{MR}$) can be controlled by the relative phase $\Phi$ of the microwaves.\cite{Phase Andre}

In addition to such coherent spin rectification effects, it is known that at the interface between a ferromagnetic and a nonmagnetic layer, microwave excitation may generate a spin polarized current flowing across the interface via the spin pumping effect\cite{RMP2005}. This effect has been observed in a few striking experiments by measuring either transmission electron spin resonance\cite{Silsbee1979} or enhanced magnetization damping\cite{Heinrich2003}. It involves FMR, exchange coupling and non-equilibrium spin diffusion.  In our opinion the physical picture of spin pumping was best explained in the classical paper of Silsbee $\mathit{et. al.}$ [Ref. \onlinecite{Silsbee1979}], which highlighted the key mechanism of dynamic exchange coupling between the precessing magnetization and the spin polarized current. Such a dynamic coupling significantly "amplifies" the effect of the rf $\textbf{h}$ field in generating non-equilibrium spins. It was later proposed that the spin current generated via spin pumping may also induce a photovoltage, either across the interface in a spin battery\cite{RMP2005, Wang spin-pumping, VanWees spin-pumping}, or within the nonmagnetic layer via the inverse spin Hall effect \cite{Saitoh ISHE, Mosendz ISHE, Mosendz ISHE2, Azevedo ISHE}. Recent experiments performed
on magnetic bilayers\cite{Liu SHE} have found that spin-pumping induced dc voltage (the term $V_{SP}$ in Table I) should be about two orders of magnitude smaller than spin Hall induced spin rectification (the term labeled $V_{SH}$).  In contrast to phase sensitive coherent spin rectification effects, the proposed spin-pumping photovoltage is based on incoherent spin diffusion and FMR absorption. Hence, the anticipated FMR line shape is symmetric and phase-independent.

\begin{widetext}
\begin{table} [h]
\begin{center}
\label{table1} \caption{Relation and distinctions between different mechanisms for microwave photovoltages induced by FMR. (For simplicity we consider only one matrix element of $\hat{\chi}$ and $\hat{\chi}_{j}$ which is responsible for the spin rectification. $\tilde{j}$ and $\tilde{m}$ denote a corresponding component of the time-dependent current and magnetization, respectively.)}
\begin{tabular}{ccccccc}
  \hline\hline
  ac driving~~~ & $\tilde{e}=e_{0}e^{-i\omega t}$~~~ & $\tilde{j}=j_{0}e^{-i\omega t}$~~~
   & $\tilde{h}=h_{0}e^{-i(\omega t-\Phi)}$~~~ & $\tilde{j}_{s}=j_{S}e^{-i\omega t}$ & ~ & ~ \\
  \hline

  Effect & Ohm's law & spin Hall &  field torque & spin torque & spin rectification & spin pumping\\
  \hline
  \\
  dc voltage &  ~ & ~ & ~ & ~ & $V\sim \langle Re(\tilde{j})\cdot Re(\tilde{m})\rangle$ & $V\sim |\tilde{m}|^{2}$ \\
  \\
  \hline\hline

  \\
  Thin film & $\tilde{j}=\sigma~\tilde{e}$  & ~ & $\tilde{m}=\chi e^{i\Theta}\tilde{h}$ & ~ & $V=V_{MR}\cdot(e_{0}h_{0})$ & ~\\
  \\
  \hline

   \\
   Bilayer & $\tilde{j}=\sigma~\tilde{e}$ & $\tilde{j}_{S}$ & ~~~$\tilde{m}=\chi e^{i\Theta}\tilde{h}$~~~+ & $\chi_{j} e^{i\vartheta}\tilde{j}_{S}$~~~ &~~ $V= V_{MR}\cdot(e_{0}h_{0})+V_{SH}\cdot(j_{0}j_{S})$ & $+~V_{SP}\cdot|m|^{2}$\\
   \\
   \hline

   \\
   MTJ & $\tilde{j},~ \tilde{j}_{S}$ & ~ & ~~~$\tilde{m}=\chi e^{i\Theta}\tilde{h}$~~~+ & $\chi_{j} e^{i\vartheta}\tilde{j}_{S}$~~~ &~~ $V= V_{MR}\cdot(e_{0}h_{0})+V_{SD}\cdot(j_{0}j_{S})$ & ~\\
   \\
  \hline\hline

\end{tabular}
\end{center}
\begin{flushleft}
\mbox{$V_{MR}$: Spin Rectification caused by \textit{MagnetoResistances};\cite{Gui SRE, Yamaguchi SRE}}\\
\mbox{$V_{SH}$: Spin Rectification caused by \textit{Spin Hall} effect;\cite{Liu SHE}}\\
\mbox{$V_{SD}$: Spin Rectification caused by \textit{Spin Diode} effect;\cite{Tulapurkar Spindiode, Kubota Spindiode, Sankey Spindiode}}\\
\mbox{$V_{SP}$: Photovoltage caused by \textit{Spin Pumping}.\cite{VanWees spin-pumping, Saitoh ISHE, Mosendz ISHE, Mosendz ISHE2, Azevedo ISHE}}
\end{flushleft}
\end{table}
\end{widetext}

From the above discussion, it is clear that the line shape analysis plays the essential role in distinguishing the microwave photovoltage generated by different mechanisms. This issue has been partially addressed by a number of theoretical\cite{Kupferschmidt2006, Kovalev2007} and experimental works\cite{Tulapurkar Spindiode, Kubota Spindiode, Sankey Spindiode} studying nanostructured MTJs where the photovoltage is dominated by the spin torque induced spin rectification. Enlightened by these works and also based on our own previous studies \cite{Nikolai2007PRB,Phase Andre}, we discuss in the following the critical issue of FMR line shape analysis in microstructured devices, where the field and spin torque induced spin rectification may have comparable strength. Our theoretical consideration and experimental data demonstrate the pivotal role of the relative phase $\Phi$, which was often under-estimated in previous studies. Via systematic studies with different device structures, measurement configurations and frequency ranges, we find that $\Phi$ has to be calibrated at different microwave frequencies for each device independently. Hence, our results are in strong contradiction with the recent experiment performed on microstructured magnetic bilayers for quantifying the spin Hall angles, where $\Phi$ was declared to be zero for all devices at different microwave frequencies\cite{Mosendz ISHE, Mosendz ISHE2}.

\section{FMR Line Shape}

\subsection{The Characteristic Signature}

From Table I, the role of the phase in the FMR line shape symmetry can be understood by considering the spin rectified voltage $V \propto
\langle Re(\tilde{j})\cdot Re(\tilde{m}) \rangle$. For spin rectification induced by the field torque, depending on the experimental configuration, at least one matrix component $\chi$ of the Polder tensor $\widehat{\chi}$ will drive the FMR; whether an on or off-diagonal component is responsible for the magnetization precession depends on the measurement configuration.  Since $\textbf{m}=\widehat{\chi}\textbf{h}$, $Re(\tilde{m}) \propto Re(\chi)\cos(\omega t - \Phi) +Im(\chi)\sin(\omega t -
\Phi)$. Therefore after time averaging a time independent dc voltage is
found $V(\Phi) \propto [Re(\chi)\cos(\Phi) - Im(\chi)\sin(\Phi)]$.  It is well known that for diagonal matrix elements, $Re(\chi)$ has a dispersive line shape while $Im(\chi)$ has a symmetric line shape.  However since the on and off-diagonal susceptibilities differ by a phase of $\pi/2$, if the FMR is driven by an off-diagonal susceptibility, the roles are reversed and $Re(\chi)$ has a symmetric line shape while $Im(\chi)$ has a dispersive line shape.

Based on the simple argument leading to the above $V(\Phi)$
expression, one can see that the line shape symmetry has a
characteristic dependence on the relative phase $\Phi$ between electric
and magnetic fields.  Thus when measuring FMR based on the field torque induced spin
rectification effect, it is important to consider the relative
phase, whereas for a spin pumping measurement which measures
$|\textbf{m}|^2$, or for a spin torque induced spin
rectification which involves
$|\textbf{j}|^2$, the relative phase does not influence the
experiment.  In the next two sections, a detailed analysis is given by solving the
Landau-Lifshitz-Gilbert equation, which leads to analytical formulae describing
the symmetric and dispersive line shapes for different measurement configurations.

\subsection{The Dynamic Susceptibility}
The Landau-Lifshitz-Gilbert equation provides a phenomenological
description of ferromagnetic dynamics based on a torque provided
by the internal magnetic field $\textbf{H}_i$ which acts on the
magnetization \textbf{M}, causing it to precess\cite{Gilbert2004}

\begin{equation}
\label{LLG}
\frac{d\textbf{M}}{dt}=-\gamma(\textbf{M}\times\textbf{H}_i)+\frac{\alpha}{M}
\left(\textbf{M} \times \frac{d\textbf{M}}{dt}\right).
\end{equation}

Here $\gamma$ is the effective electron gyromagnetic ratio and
$\alpha$ is the Gilbert damping parameter which can be used to
determine the FMR line width $\Delta H$, according to $\Delta H
\sim \alpha \omega/\gamma$.  For the case of microwave induced
ferromagnetic resonance Eq. (\ref{LLG}) can be solved by splitting
the internal field into dc and rf components and taking the
applied dc field \textbf{H}, along the $z$-axis.  We can relate the
internal field $\textbf{H}_i = \textbf{H}_{0i}+\textbf{h}_ie^{-i
\omega t}$, to the applied field through the demagnetization
factors $N_k$, $H_{0iz} = H-N_zM_0$, $h_{ik} =
h_ke^{i\Phi_k}-N_km_k$, where $\Phi_k$ is the relative phase shift
between the electric and magnetic fields in the $k^{th}$ direction
and $\textbf{M}_0$ is the dc magnetization also along the $z$-axis.
With the magnetization separated into dc and rf contributions
$\textbf{M}=\textbf{M}_0+\textbf{m}e^{-i\omega
t}$, the solution of Eq. (\ref{LLG})
yields the dynamic susceptibility tensor $\widehat{\chi}$ which
relates the magnetization \textbf{m} to the externally applied rf
field \textbf{h}
\begin{align}
\label{chifinal}
\nonumber
\textbf{m}=\widehat{\chi}\textbf{h} &= \left(
\begin{array}{ccc}
\chi_{xx} & i\chi_{xy} & 0 \\
-i\chi_{xy} & \chi_{yy} & 0 \\
0 & 0 & 0
\end{array}
\right) \textbf{h} \\
& = \left(
\begin{array}{ccc}
|\chi_{xx}| & |\chi_{xy}|e^{i\frac{\pi}{2}} & 0 \\
|\chi_{xy}|e^{-i\frac{\pi}{2}} & |\chi_{yy}| & 0 \\
0 & 0 & 0
\end{array}
\right) \textbf{h}e^{i\Theta},
\end{align}
where $\Theta = \arctan[\Delta H/(H-H_r)]$ is the spin resonance
phase\cite{Phase Andre} which describes the phase shift between the response and the
driving force in terms of the line width $\Delta H$ and the
resonance field $H_r$ which are constant for a fixed frequency.
$\Theta$ will change from 180$^\circ$ (driving force out of phase)
to 0$^\circ$ (driving force in phase) around the resonance
position, in a range on the order of $\Delta H$, passing through
90$^\circ$ at resonance.  This represents the universal feature of
a resonance; the phase of the dynamic response always lags behind the driving
force.\cite{Landau1969}

To emphasize the resonant feature of the susceptibility tensor
elements we define the symmetric Lorentz line shape $L$, and the
dispersive line shape $D$ as
\begin{align}
\nonumber
L &= \frac{\Delta H^2}{(H-H_r)^2+\Delta H^2}, \\
D &= \frac{\Delta H(H-H_r)}{(H-H_r)^2+\Delta H^2}.
\end{align}

Clearly the spin resonance phase can also be written in terms of
$L$ and $D$ as $\Theta = \arctan[\Delta H/(H-H_r)] = \arctan(L/D)$
so that $L \propto \sin(\Theta)$ and $D \propto \cos(\Theta)$.
Therefore $L$ and $D$ carry the resonant information of the
susceptibility tensor.

Using $L$ and $D$ allows the elements of $\widehat{\chi}$ to be
written as $(\chi_{xx}, \chi_{xy}, \chi_{yy}) =
\left(D+iL\right)(A_{xx}, A_{xy}, A_{yy})$.   $A_{xx}, A_{xy}$ and
$A_{yy}$ are real amplitudes which are related to the sample
properties

\begin{align}
\label{As}
\nonumber
A_{xx} &= \frac{\gamma M_0(M_0N_y+(H-N_zM_0))}{\alpha\omega(2(H-N_zM_0)+M_0(N_z+N_y))},\\
\nonumber
A_{xy} &= -\frac{M_0}{\alpha(2(H-N_zM_0)+M_0(N_z+N_y))},\\
A_{yy} &= \frac{\gamma M_0(M_0N_x+(H-N_zM_0))}{\alpha\omega(2(H-N_zM_0)+M_0(N_z+N_y))}.
\end{align}

Since these amplitudes are real all components of $\widehat{\chi}$ include both a dispersive and a Lorentz line shape determined solely from the $D+iL$ term. However, in a transmission experiment performed using a resonance cavity $|m|^2 \propto L^2 +D^2 = L$ is
measured.  This product removes the phase dependence carried by $L$ and $D$ and leaves only the Lorentz line shape. For the same reason, the microwave photovoltage induced by spin pumping (the $V_{SP}$ term in Table I) has a symmetric line shape.

The susceptibility for the two cases of in-plane and
perpendicularly applied dc magnetic fields can easily be found
from Eq. (\ref{As}) by using the appropriate demagnetization
factors.  When the lateral dimensions are much larger than the
thickness, N$_x$ = N$_z$ = 0 and N$_y$ = 1 for an in-plane field
and N$_x$ = N$_y$ = 0 and N$_z$ = 1 for a field applied at a small
angle from the perpendicular.  In this paper, we focus on the in-plane case. The line shape analysis for the perpendicular case can be found in Ref. \onlinecite{Phase Andre}. In both cases the form of the susceptibility,
$\chi \propto  D+iL$, describes the ferromagnetic resonance
line shape where each element of $\widehat{\chi}$ is the sum of an
antisymmetric and symmetric Lorentz line shape. As we describe in the next section, via the $V_{MR}$ term of the spin
rectification effect, the symmetry properties of the dynamic
susceptibility influence the symmetry of the electrically detected
FMR which can be controlled by tuning the relative
electromagnetic phase $\Phi$.

\subsection{Spin Rectification Induced by the Field Torque}

The field-torque spin rectification effect results in the production of a dc
voltage from the non-linear coupling of rf electric and magnetic
fields. For example, it may follow from the generalized Ohm's
law\cite{Juretschke, Ohm's-law}
\begin{equation}
\label{gohm}
\textbf{J}=\sigma\textbf{E}_{0} - \frac{\sigma\Delta\rho}{\textbf{M}^2}(\textbf{J}\cdot\textbf{M})
\textbf{M}+\sigma R_{H}\textbf{J}\times\textbf{M},
\end{equation}
where $\sigma$ is the conductivity, $\Delta \rho$ is the resistivity change due to AMR and $R_H$ is the extraordinary Hall coefficient.

\begin{figure} [t]
\centering \epsfig{file=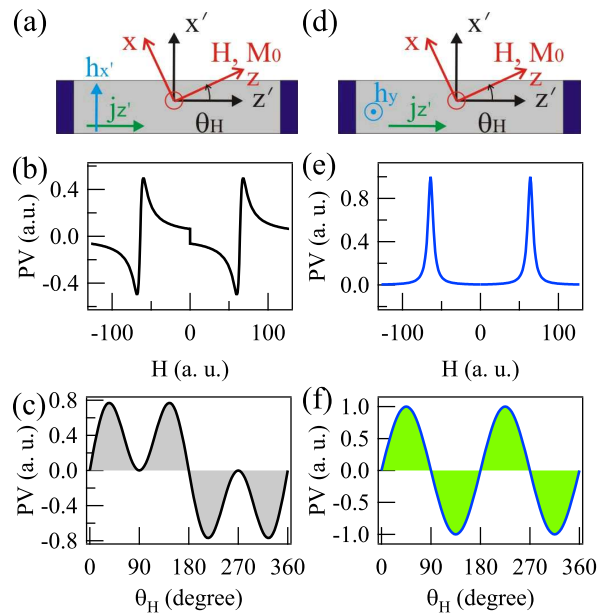,width=8 cm} \caption{(color online). Left panel (a) Coordinate system for an in-plane dc $H$ field applied along the $z$-axis at an angle $\theta_H$ with respect to the $z^\prime$-axis, with a rf $h$-field along the $x^\prime$-axis. (b) The calculated photovoltage (PV) spectrum at $\theta_H=45^\circ$ and (c) the calculated amplitude of the PV spectrum at FMR as a function of $\theta_H$ according to Eq. (\ref{Vx}). Right Panel (d)-(f) are the same as (a)-(c), respectively, but with a rf $h$-field along the $y$-axis, and calculations are according to Eq. (\ref{Vy}). In both cases, $\Phi$ is assumed to be zero.} \label{fig2}
\end{figure}

As shown in Fig. \ref{fig2}, we use two coordinate systems to describe a long narrow strip under the rotating in-plane magnetic field $\textbf{H}$. The sample coordinate system $(\widehat{\textbf{x}}^\prime, \widehat{\textbf{y}},
\widehat{\textbf{z}}^\prime)$ is fixed with the sample length along the $z^\prime$ direction
and the sample width in the $x^\prime$ direction. The measurement coordinate system $(\widehat{\textbf{x}}, \widehat{\textbf{y}},
\widehat{\textbf{z}})$ rotates with the $\textbf{H}$ direction which is along the $\widehat{\textbf{z}}$ axis. We define $\theta_H$ as the angle between the direction of the strip and the in-plane applied static magnetic field
(\textit{i.e.}, between the $z^\prime$ and $z$ directions). In both coordinate systems, the $\widehat{\textbf{y}}$ axis is along the normal of the sample plane. In the case of a sample length much larger than the width, the rf
current, $\tilde{j}=j_{z^\prime}e^{-i\omega t}$ flows along the
strip direction $z^\prime$.  In this geometry the field due to the
Hall effect will only be in the transverse direction and will not
generate a voltage along the strip. Taking the
time average of the electric field integrated along the $z^\prime$
direction, the photovoltage is found as\cite{Gui SRE, Nikolai2007PRB}
\begin{equation}
\label{V1} V=\frac{\Delta R}{M_0}\langle Re(\tilde{j})\cdot Re(\tilde{m}_x)\rangle\sin(2\theta_H),
\end{equation}
where $\Delta R$ is the resistance change due to the AMR effect
and the $\sin(2\theta_H)$ term is a result of the AMR effect which
couples \textbf{J} and \textbf{M}.


The susceptibility tensor given by Eqs. (\ref{chifinal}) and
(\ref{As}) can be used to write $\tilde{m}_x$ in terms of the rf
$\textbf{h}$ field.  Since $\textbf{M}_0$ and $\textbf{H}$ are
both along the $z$-axis, only the components of \textbf{h}
perpendicular to \textbf{z} will contribute to \textbf{m}.
However, since the rf current flows in
the $z^\prime$ direction, to calculate the rectified voltage, $\tilde{m}_x$ must be transformed into the $(x^\prime, y, z^\prime)$
coordinate system by using the rotation
$(\widehat{\textbf{x}}, \widehat{\textbf{y}},
\widehat{\textbf{z}})=(\cos(\theta_H)\widehat{\textbf{x}}^\prime-\sin(\theta_H)\widehat{\textbf{z}}^\prime,~
\widehat{\textbf{y}},~
\sin(\theta_H)\widehat{\textbf{x}}^\prime+\cos(\theta_H)\widehat{\textbf{z}}^\prime)$, which introduces an additional $\theta_H$ dependence
into the photovoltage.  We find that the photovoltage can be written in terms of the
symmetric and antisymmetric Lorentz line shapes, $L$ and $D$, as

\begin{equation}
\label{fullV}
V=\frac{\Delta R}{2M_0}j_{z^\prime}\left(A_LL + A_DD \right),
\end{equation}
where
\begin{align}
\label{LD}
\nonumber
A_L &= \sin(2\theta_H)[-A_{xx}h_{x^\prime}\cos(\theta_H)\sin(\Phi_{x^\prime}) \\
\nonumber
& \qquad -A_{xy}h_{y}\cos(\Phi_{y})+A_{xx}h_{z^\prime}\sin(\theta_H)\sin(\Phi_{z^\prime})], \\
\nonumber
A_D &=\sin(2\theta_H)[A_{xx}h_{x^\prime}\cos(\theta_H)\cos(\Phi_{x^\prime}) \\
& \qquad  -A_{xy}h_{y}\sin(\Phi_{y})- A_{xx}h_{z^\prime}\sin(\theta_H)\cos(\Phi_{z^\prime})],
\end{align}
and $\Phi_{x^\prime}, \Phi_{y}$ and $\Phi_{z^\prime}$ are the
relative phases between electric and magnetic fields in the
${x^\prime}, {y}$ and ${z^\prime}$ directions, respectively.

The amplitudes of the Lorentz and dispersive line shape
contributions show a complex dependence on the relative phases for
the $x^\prime, y$ and $z^\prime$ directions and in general both
line shapes will be present.  However, depending on the
experimental conditions, this dependence may be simplified.  For
instance when $h_{x^\prime}$ is the dominate driving field as shown in Fig. \ref{fig2}(a), we may
take $h_{y}$ = $h_{z^\prime} \approx 0$ and $\Phi_{x^\prime} = \Phi$, which results in
\begin{align}
\label{Vx}
\nonumber
V=-\frac{\Delta R}{2M_0}j_{z^\prime}&A_{xx}h_{x^\prime}\cos(\theta_H)\sin(2\theta_H) \\
& \left[L\sin(\Phi) - D \cos(\Phi)\right].
\end{align}
From Eq. (\ref{Vx}) we see that the photovoltage line shape
changes from purely symmetric to purely antisymmetric in $90^\circ$
intervals of $\Phi$, being purely antisymmetric when
$\Phi = n \times 180^\circ$ and purely symmetric when
$\Phi = (2n+1) \times 90^\circ,~n=0, \pm~1, \pm~2 \dots$.

As shown in Fig. \ref{fig2}(b) and (c), the photovoltage in Eq. (\ref{Vx}) also shows symmetries depending
on the static field direction $\theta_H$.  Since \textbf{H} $\to$
-\textbf{H} corresponds to $\theta_H \to \theta_H+180^\circ$,
$V(H)=-V(-H)$.  Furthermore at $\theta_H =
n \times 90^\circ,~n=0, \pm~1, \pm~2 \dots$ the voltage will be zero.

Similarly when $h_{y}$ dominates as shown in Fig. \ref{fig2}(c), we take $h_{x^\prime}$ =
$h_{z^\prime}\approx 0$ and $\Phi_{y} = \Phi$ which results in a voltage
\begin{align}
\label{Vy}
\nonumber
V=-\frac{\Delta R}{2M_0}j_{z^\prime}&A_{xy}h_{y}\sin(2\theta_H) \\
& \left[L \cos(\Phi) + D\sin(\Phi)\right].
\end{align}
The symmetry properties are now such that the line shape is purely
symmetric when $\Phi = n \times 180^\circ$ and purely antisymmetric when $\Phi =
(2n+1) \times 90^\circ,~n=0, \pm~1, \pm~2 \dots$.  Also the photovoltage
determined by Eq. (\ref{Vy}) is now symmetric with respect to $H$
under $\theta_H \to \theta_H +180^\circ$ so that $V(H)=V(-H)$ as
shown in Fig. \ref{fig2}(e). Therefore, experimentally the
different symmetry of the FMR at $H$ and $-H$ can be
used as an indication of which component of the \textbf{h} field is
dominant.

Both Eq. (\ref{Vx}) and Eq. (\ref{Vy}) demonstrate that a change
in the relative electromagnetic phase is expected to result in a
change in the line shape of the electrically detected FMR.  It is
worth noting that when the relative phase $\Phi = 0$, the line
shape is purely antisymmetric for FMR driven by $h_{x^\prime}$ and
purely symmetric for FMR driven by $h_y$ as illustrated in Fig.
\ref{fig2}(b) and \ref{fig2}(e), respectively. In the general case
when $\tilde{m}_x$ is driven by multiple $\textbf{h}$ components, Eq.
(\ref{fullV}) must be used in combination with angular
($\theta_H$) dependent measurements in order to distinguish
different contributions.

\subsection{The Physics of $\Phi$}

It is clear therefore that for field torque induced spin rectification, the relative phase
$\Phi$ between the microwave electric and magnetic fields plays the pivotal role in the FMR line shape. Note that $\Phi$ is a material and frequency dependent property which is related to the losses in the
system.\cite{Jackson, Heinrich1990, Heinrich1993} When a plane
electromagnetic wave propagates through free space the electric
and magnetic fields are in phase and orthogonal to each
other.\cite{Born1999} However when the same electromagnetic wave travels
through a dispersive medium where the wave vector is complex, the
imaginary contribution can create a phase shift between electric
and magnetic fields.  The most well known example is that of a
plane electromagnetic wave moving in a conductor \cite{Jackson} where Faraday's
law gives a simple relation between electric and magnetic fields,
$\omega \mu \textbf{h} = \textbf{k} \times \textbf{e}$.  Therefore
the complex part of the wave vector \textbf{k} will induce a phase
shift between electric and magnetic fields.  Although the field
will exponentially decay inside a conductor, it will still
penetrate a distance on the order of the skin depth, and in a
perfect conductor the conductivity, which produces an imaginary
dielectric constant, will result in a phase shift of $45^\circ$
between the electric and magnetic fields.\cite{Jackson}

In a complex system such as an experimental set up involving waveguides, coaxial
cables, bonding wires and a sample holder, which are required for electrical
FMR detection, the relative phase cannot be simply calculated.
Nevertheless losses in the system which can be characterized in a
variety of ways, such as through the wave
impedance,\cite{Heinrich1990, Heinrich1993} will lead to a phase
shift between electric and magnetic fields which will influence
the FMR line shape.

Although the physics of $\Phi$ is in principle contained in Maxwell's equations, due to the lack of technical tools for simultaneously and coherently probing both $\textbf{e}$ and $\textbf{h}$ fields, the effect of the relative phase had often been ignored until the recent development of spintronic Michelson interferometry\cite{Phase Andre}. In the following we provide systematically measured data showing the influence of the relative phase $\Phi$
on the line shape of FMR which is driven by different $\textbf{h}$ field components.

\section{Experimental Line Shape Measurements}

\subsection{\textbf{h}$_\textbf{y}$ Dominant FMR}

\begin{figure} [t]
\centering \epsfig{file=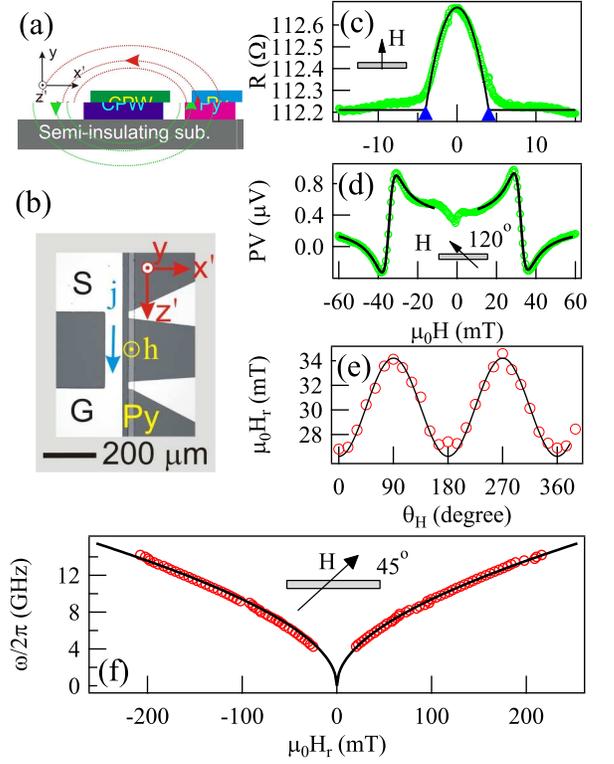,width=8 cm} \caption{(color
online). (a) Schematic diagram of the first generation spin dynamo where the Py strip is located beside the CPW.  The dominate
magnetic field in the Py is the Oersted field in the $y$ direction
due to the current in the CPW.  (b) Micrograph of the
device. (c) Magnetoresistance at $\theta_H = 90^\circ$.  AMR is
seen to be $\sim 0.4 \%$.  Arrows denote the anisotropic field,
$\mu_0H_A$ = 4.0 mT.  Open circles are experimental data and solid
curve is the fitting result using $R(0) = 112.66 ~\Omega, \Delta R
= 0.47 ~\Omega, H_A = 4.0$ mT. (d) Electrically detected FMR at
$\theta_H = 120^\circ$ and $\omega/2\pi$ = 5 GHz showing an almost purely
dispersive line shape ($\Phi \simeq 90^\circ$). Fit is according to Eq.
(\ref{Vy}) with $\mu_0 \Delta H$ = 3.6 mT, $\mu_0 H_r$ = 32.2 mT.
(e) Oscillating $H_r$ dependence on the static field direction
$\theta_H$ with amplitude $2H_A$.  (f) Dependence of FMR frequency
on the resonant field $H_r$ at $\theta_H = 45^\circ$.  Open
circles are experimental data and the solid line is the fit
according to $\omega = \gamma \sqrt{|H_r|(|H_r|+M_0)}$.}
\label{fig3}
\end{figure}

In order to use the $h_{y}$ field to drive FMR a first generation
spin dynamo was used where a Cu/Cr coplanar waveguide (CPW) was
fabricated beside a Py microstrip
with dimension 300 $\mu$m $\times$ 20 $\mu$m $\times$ 50 nm on a
SiO$_2$/Si substrate as shown in Fig. \ref{fig3}(a).  A microwave
current is directly injected into the CPW and flows in the
$z^\prime$ direction inducing a current in the Py strip also along
the $z^\prime$-axis.  In this geometry the dominant rf
$\textbf{h}$ field in the Py will be the Oersted field in the $y$
direction produced according to Amp\`{e}re's Law.  This field will
induce FMR precession with the same cone angle independent of the
static \textbf{H} orientation.

The AMR resistance depends on the orientation
of the magnetization relative to the current and follows the relation $R(H)=R(0)-\Delta R
\sin^2(\theta_M)$, where $\theta_M$ (not shown) is the angle
between the magnetization and the current direction. For Py the AMR effect, which is responsible for the spin rectification, is observed to produce a
resistance change of $\Delta R/R(0) \sim 0.4$ \%.  When
\textbf{H} is applied along the $x^\prime$-axis, \textit{i.e.}, the in-plane hard
axis, the magnetization \textbf{M} tends to align toward the
static field \textbf{H} and the angle $\theta_M$ is determined by $\sin(\theta_M)=H/H_A$ for $H<H_A$, where $H_A = N_{x^\prime}M_0$ is the in-plane shape anisotropy field. The measured data (symbols) shown in Fig.
\ref{fig3}(c) is fit (solid curve) according to $R(H)=R(0)-\Delta R \sin^2(\theta_M)$ with $R(0)
= 112.66~\Omega$, $\Delta R = 0.47~\Omega$, $\mu_0H_A=4.0$ mT, and $N_{x^\prime}$=0.004.

Fig. \ref{fig3}(d) shows that the line shape at $\theta_H =
120^\circ$ and $\omega/2\pi$ = 5 GHz is almost purely dispersive, indicating that at this frequency $\Phi
\sim 90^\circ$ according to Eq. (\ref{Vy}).  The $\theta_H$
dependence of $H_r$ is shown in Fig. \ref{fig3}(e) and can be well
fit by the function $\omega = \gamma
\sqrt{(|H_r|+H_A\cos(2\theta_H))[|H_r|+M_0-H_A(1+\sin^2(\theta_H)]}$
by taking the shape anisotropy field $H_A$ along the $x^\prime$-axis into account.\cite{FMR1966}  As expected the amplitude of
these oscillations is $\mu_0H_A$ = 4.0 mT.  The frequency
dependence of $H_r$ at $\theta_H=45^\circ$ is shown in Fig. \ref{fig3}(f) and is fit
using $\omega = \gamma \sqrt{|H_r|(|H_r|+M_0)}$ with $\gamma/2\pi
= 29.0 ~ \mu_0$GHz/T and $\mu_0 M$=1.0 T.

\begin{figure} [t]
\centering \epsfig{file=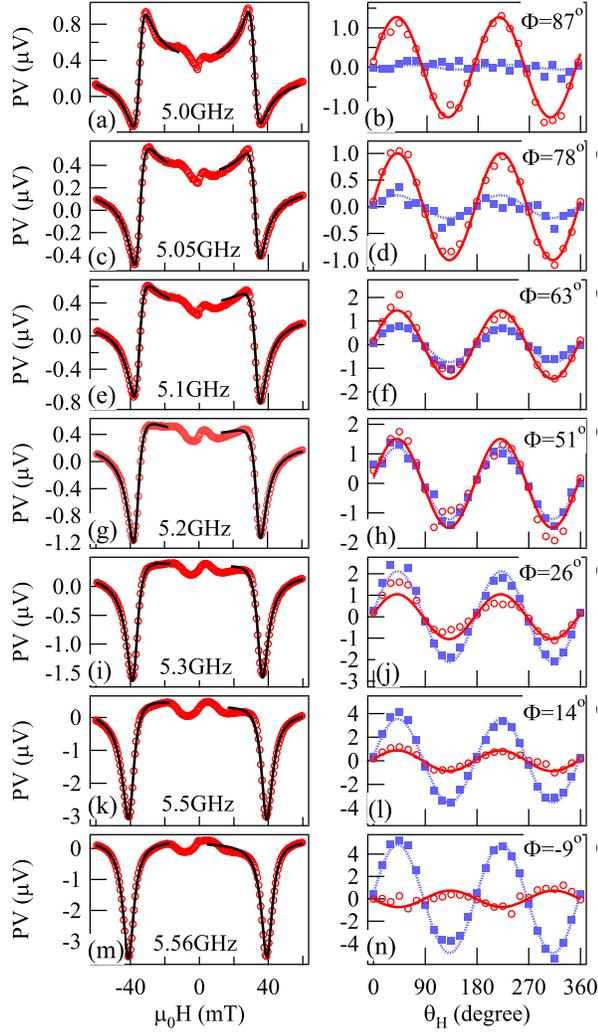,width=8 cm} \caption{(color
online). Data shown for a first generation spin dynamo.  FMR spectra at
$\theta_H=120^\circ$ for several frequencies from 5.0 to 5.56 GHz
with corresponding Lorentz and dispersive amplitudes as a function
of $\theta_H$. Circles and squares indicate the Lorentz and
dispersive amplitudes of Eq. (\ref{Vy}) respectively and show a
$\sin(2\theta_H)$ dependence as expected.  Solid and dashed curves
are $\sin(2\theta_H)$ functions.}  \label{fig4}
\end{figure}

\begin{figure} [t]
\centering \epsfig{file=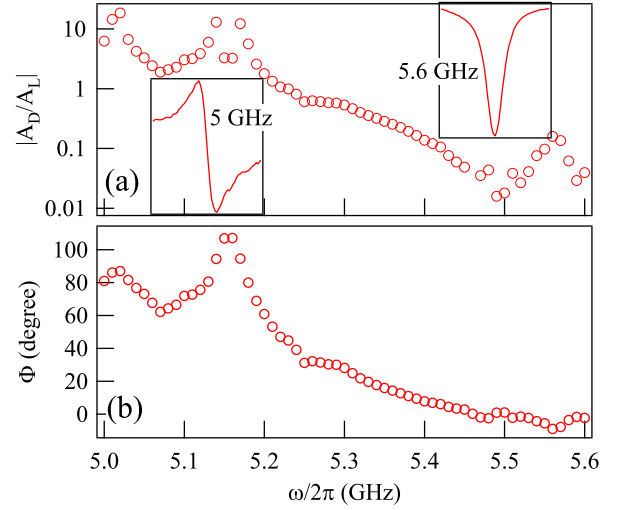,width=8 cm} \caption{(color
online). (a) The $A_D/A_L$ ratio as a function of $\omega/2\pi$
showing the line shape change from dispersive at 5 GHz (left
inset) to Lorentz at 5.6 GHz (right inset) with a step size of
0.01 GHz. (b) $\Phi$ dependence on $\omega/2\pi$ over same
frequency interval showing the same dependence as $A_D/A_L$.}
\label{fig5}
\end{figure}

By systematically measuring the line shape as a function of the
microwave frequency, we observe the interesting results of Fig.
\ref{fig4}.  The FMR line shape is observed to change from almost
purely dispersive at $\omega/2\pi$ = 5 GHz to almost purely
symmetric at $\omega/2\pi$ = 5.56 GHz.  As discussed before, the
line shape may be affected by the \textbf{h} orientation, $i.e.$, different \textbf{h} vector components will affect the line
shape differently. Hence, if changing the microwave frequency
changes the dominant driving field, the line shape may change. To
rule out such a possibility an angular dependent
experiment was performed to measure the line shape at different
$\theta_H$ for each frequency $\omega$. The results are plotted on the right
panel of Fig. \ref{fig4} which shows the sinusoidal curves for
the Lorentz, $A_L$, and dispersive, $A_D$, amplitudes (dashed and
solid curves respectively) as a function of the static field
angle $\theta_H$. Both the Lorentz and dispersive amplitudes are found to follow a
$\sin(2\theta_H)$ dependence on the field angle in agreement with
Eq. (\ref{Vy}) indicating that the magnetization precession is
indeed dominantly driven by the $h_{y}$ field. Therefore the line shape change indicates that the relative phase $\Phi$ is frequency dependent.
As shown in Fig. \ref{fig5}(a), at $\omega/2\pi$ = 5 GHz the amplitude of $A_D$
is approximately one order of magnitude larger than $A_L$, while
at $\omega/2\pi$ = 5.56 GHz $A_D$ is one order of magnitude less
than $A_L$.  Such a large change in $A_L/A_D$ shows that in a
microwave frequency range as narrow as 0.6 GHz, the relative phase
$\Phi$ can change by 90$^\circ$. Fig. \ref{fig5}(b) shows $\Phi$ determined by using Eq. (\ref{Vy}), which smoothly changes with microwave
frequency except for a feature near 5.18 GHz, which is possibly caused by a resonant waveguide mode at this frequency.

Such a large change of $\Phi$ within a very narrow range of microwave frequency indicates the complexity of wave physics.  Note that microwaves at $\sim$ 5 GHz have wavelengths on the order of
a few centimeters which are much larger than the sub-millimeter
sample dimensions.  Consequently the microwave propagation depends
strongly on the boundary conditions of Maxwell's equations which
physically include the bonding wire, chip carrier, as well as the
sample holder. This is similar to the microwave propagation in a waveguide where
the field distribution $i.e.$ the waveguide modes, are known to
depend strongly on boundary conditions and
frequency.\cite{Guru2004} Despite the complex wave properties, the key message of our results is clear and consistent with the consideration of the physics of the relative phase: it shows that in order to properly analyze the FMR line shape, $\Phi$ has to be determined for each frequency independently.


\begin{figure} [t]
\centering \epsfig{file=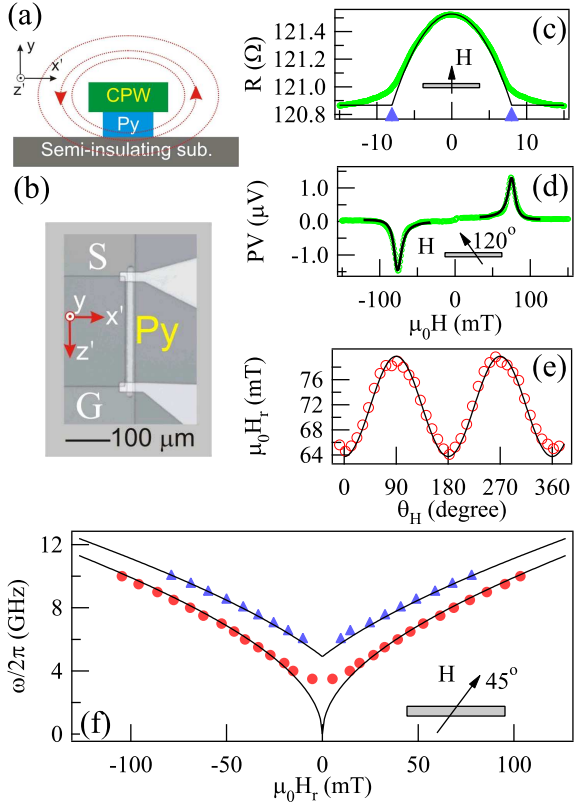,width=8 cm} \caption{(color
online). (a) Schematic diagram of the second generation spin dynamo where the Py strip is located underneath the CPW.  In this
case the dominant magnetic field in the Py is the Oersted field in
the $x^\prime$ direction due to the field in the CPW.  (b)
Micrograph of the Py CPW device. (c) Magnetoresistance at
$\theta_H = 90^\circ$.  AMR is seen to be $\sim 0.5 \%$.  Arrows
denote the anisotropic field, $\mu_0H_A$ = 8.0 mT.  Open circles
are experimental data and solid curve is the fitting result using
$R(0) = 121.53 ~\Omega$ and $\Delta R = 0.66 ~\Omega$. (d)
Electrically detected FMR at $\theta_H = 120^\circ$ and
$\omega/2\pi = 8$ GHz showing a nearly symmetric Lorentz line shape. Fit
is according to Eq. (\ref{Vy}) with $\mu_0 \Delta H = 6.0$ mT ,
$\mu_0 H_r = 76.5$ mT and $\Phi = -102^\circ$. (e) Oscillating
$H_r$ dependence on the static field direction $\theta_H$ with
amplitude $2H_A$.  (f) Dependence of FMR frequency on the resonant
field $H_r$ at $\theta_H = 45^\circ$.  Solid circles show the FMR
frequency dependence while the solid triangles are the standing
SWR frequency dependence.  The solid line is a fit to $\omega =
\gamma \sqrt{|H_r|(|H_r|+M_0)}$.} \label{fig6}
\end{figure}

\subsection{\textbf{h}$_{\textbf{x}^\prime}$ Dominant FMR}

In order to drive the FMR using the rf field in the $x^\prime$
direction, $h_{x^\prime}$, a second generation spin dynamo was
fabricated with the Py strip underneath the CPW as shown in Fig.
\ref{fig6}.  In this case the 300 $\mu$m $\times$ 7 $\mu$m
$\times$ 100 nm Py strip is underneath the Cu/Cr coplanar
waveguide which is fabricated on a SiO$_2$/Si substrate.  Again a
microwave current is directly injected into the CPW and induces a
current in the $z^\prime$ direction in the Py strip.  The dominant
rf field in the Py is still the Oersted field, but due to the new
geometry it is in the $x^\prime$ direction.

Due to the smaller
width and larger thickness, the demagnetization factor,
$N_{x^\prime} = 0.008$ is twice that in the first generation
sample.  This corresponds to $\mu_0H_A$ = 8.0 mT as indicated by
the broader AMR curve in Fig. \ref{fig6}(c).  This value is
further confirmed by the $H_r$ vs $\theta_H$ plot shown in Fig.
\ref{fig6}(e). Fig. \ref{fig6}(f) shows the frequency dependence of $H_r$ for FMR
(circles) and for the first perpendicular standing spin wave
resonance (SWR) (triangles) measured at $\theta_H=45^\circ$.  The frequency dependence of $H_r$
follows $\omega = \gamma \sqrt{(|H_r|+H_{ex})(|H_r|+M_0+H_{ex})}$
where $H_{ex}$ is the exchange field.  In Fig. \ref{fig6}(f) the
standing SWR is fit using $\gamma/2\pi = 29.0~\mu_0$GHz/T ,
$\mu_0H_{ex} = 30$ mT and $\mu_0M_0 = 1.0$ T.

Similar to the results presented in the previous section, the line shape of FMR measured on the second generation sample is also found to be frequency dependent (not shown). Hence, $\Phi$ is found to be non-zero in the general case. For example, at $\omega$/2$\pi$ = 8 GHz, the line shape is found to be nearly symmetric, as shown in Fig. \ref{fig6}(d) for the FMR measured at $\theta_H=120^\circ$, which indicates $\Phi$ is close to $-90^\circ$ at this frequency. Note that our result is in direct contrast with the recent study of Ref. \onlinecite{Mosendz ISHE} and \onlinecite{Mosendz
ISHE2}, where experiments were measured in the same configuration and where it was suggested that $\Phi$ = 0$^\circ$ for all samples at all frequencies.

\begin{figure} [t]
\centering \epsfig{file=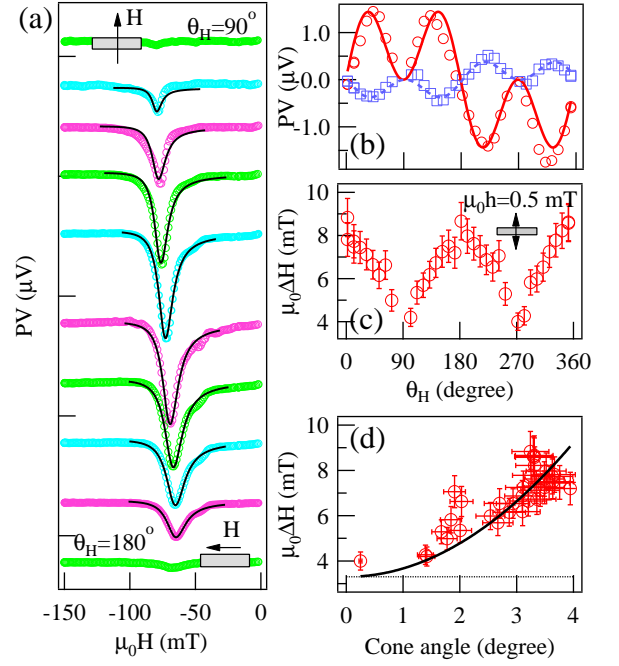,width=8 cm} \caption{(color
online). Data shown for a second generation spin dynamo.  (a) FMR line
shape at fixed frequency, $\omega/2\pi = 8$ GHz for several
$\theta_H$ from $90^\circ$ to $180^\circ$ in steps of $10^\circ$.
Open circles are experimental data and solid lines are fits using
Eq. (\ref{Vx}) with $\Phi= -102^\circ$ fixed.  (b) $A_D$ and $A_L$
shown in squares and circles respectively as a function of
$\theta_H$.  Fitting curves are $\sin(2\theta_H)\cos(\theta_H)$
functions. (c) $\Delta H$ for several values of $\theta_H$ showing
an oscillation with $\theta_H$. (d) Non-linear dependence of line
width $\Delta H$ on the cone angle.  Dashed line is the expected
linear Gilbert damping whereas the data follows the quadratic
dependence shown by the solid line.}  \label{fig7}
\end{figure}

While the line shape and hence the relative phase is found to be
frequency dependent, $\Phi$ is expected to be independent of the
static field direction $\theta_H$.  This is confirmed in Fig.
\ref{fig7}(a) which shows the line shape measured at several
values of $\theta_H$ in 10$^\circ$ increments.  The data can be
fit well using Eq. (\ref{Vx}) with a constant $\Phi = -102^\circ$
for all $\theta_H$.  It confirms that the FMR is driven by a
single \textbf{h} component, in this case the $h_{x^\prime}$
field, and that $\Phi$ does not depend on $\theta_H$. In Fig.
\ref{fig7}(b) the $\theta_H$ dependence of $A_L$ and $A_D$
(solid/circles and dashed/squares respectively) is shown.  The
circles and squares are experimental data while the solid and
dashed lines are fitting results using a
$\sin(2\theta_H)\cos(\theta_H)$ function according to Eq.
(\ref{Vx}). It provides further proof that the $h_{x^\prime}$
field is responsible for driving the FMR in this sample.

While the results from both the 1st and 2nd generation spin
dynamos show consistently that $\Phi$ is sample and frequency
dependent, the 2nd generation spin dynamos exhibit special
features in comparison with the 1st generation spin dynamos: the
reduced separation between the Py strip and CPW enhances the
$h_{x^\prime}$ field so that the line width $\Delta H$ is enhanced
by non-linear magnetization damping\cite{Gui Damping1, Gui
Damping2, Slavin nonlinear damping}, which depends on the cone
angle $\theta$ of the precession via the relation $\theta \sim
h_{x^\prime}\cos(\theta_H)/\Delta H(\theta)$. As shown in Fig.
\ref{fig7}(c), $\Delta H$ is found to oscillate between 4.0 and 9.0
mT as $\theta_H$ changes. At $\theta_H = 0^\circ$, $\theta \sim
h_{x^\prime}/\Delta H$ and the cone angle is at its largest (about
4$^\circ$). As $\theta_H$ increases from $0^\circ$ and moves
toward 90$^\circ$, $\theta$ decreases so that the non-linear
damping contribution to $\Delta H$ decreases. Using the cone angle
calculated from Fig. \ref{fig7}(c), we plot in Fig. \ref{fig7}(d)
$\Delta H(\theta)$ as a function of the cone angle. It shows that $\Delta
H$ has a quadratic dependence on the precession cone angle, which
is in agreement with our previous study in the perpendicular
$\textbf{H}$-field configuration\cite{Gui Damping1, Gui Damping2}.
We note that for cone angles above only a few degrees, the
non-linear damping already dominates the contribution to $\Delta
H$. Again, this is in direct contrast with the result of Mosendz
\textit{et al.},\cite{Mosendz ISHE,Mosendz ISHE2}, where $\Delta H$ was
found to be constant by varying $\theta_H$, indicating no influence
from non-linear damping, but the cone angle $\theta$ was estimated to be as high as 15$^\circ$ based on the line shape analysis assuming relative phase $\Phi$ = 0.

\subsection{Arbitrary \textbf{h} Vector}

Next we consider the most general case which is described by Eq.
(\ref{fullV}) where all components of \textbf{h} may contribute to
the FMR line shape.  The sample used here is a single Py strip
where a waveguide with a horn antennae provided both the electric and magnetic driving fields. The sample chip is
mounted near the centre, at the end of a rectangular waveguide and
the Py strip is directed along the short axis of the waveguide.

\begin{figure} [t]
\centering \epsfig{file=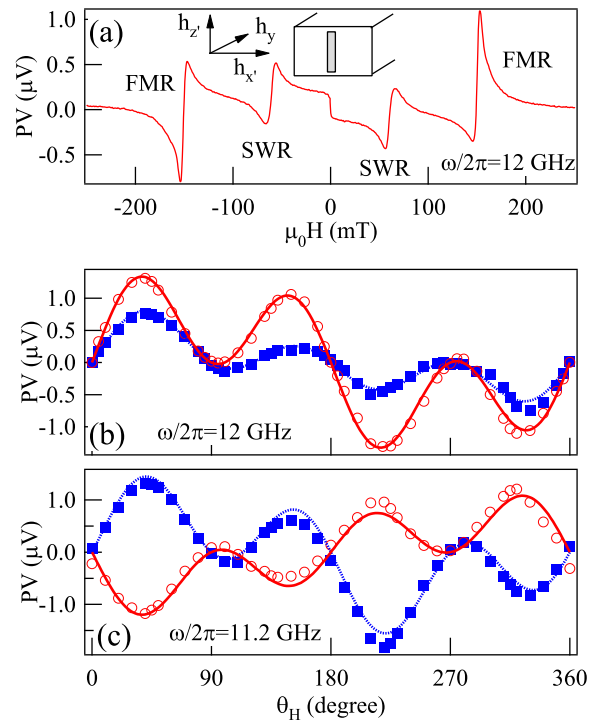,width=8 cm} \caption{(color
online). Data shown for a single Py strip with precession driven
by horn antennae field.  The strip dimensions are 3 mm $\times
~50~\mu$m $\times~ 45$ nm.  (a) Spectra showing distinct
resonances due to FMR  and SWR at $\omega/2\pi$ = 12 GHz.  (b)
Separated Lorentz and dispersive line shapes (circles and squares
respectively) as a function of $\theta_H$ from a fit to Eq.
(\ref{fullV}) at $\omega/2\pi = 12$ GHz and (c) $\omega/2\pi$ =
11.2 GHz.}  \label{fig8}
\end{figure}

In a waveguide, the electromagnetic fields are well known and in
general three components, $h_{x^\prime}, h_{y}$ and $h_{z^\prime}$
exist.\cite{Guru2004}  Figure \ref{fig8}(a) shows both the FMR and
perpendicular standing SWR at $\theta_H$ = 45$^\circ$.  Indeed
both the amplitude and the line shape are different for the two
FMR peaks located at $H$ and $-H$, which indicates the existence of
multiple \textbf{h} field components and Eq. (\ref{fullV}) and Eq.
(\ref{LD}) are needed to separate them.

This separation is done using the Lorentz and dispersive
amplitudes determined from a fit to the FMR which are plotted as a
function of $\theta_H$ in Fig. \ref{fig8}(b) and (c) for
$\omega/2\pi = 12$ and 11.2 GHz, respectively.  A fit using Eq.
(\ref{LD}) allows a separation of the contributions from each of
the $h_{x^\prime}, h_{y}$ and $h_{z^\prime}$ fields based on the
their different contributions to the $\theta_H$ dependence of the
line shape.

\begin{table} [h]
\centering
\caption{Angular separation of \textbf{h} field components for 12 and 11.2 GHz.}
\begin{tabular}{ c | c | c   }
   & 12 GHz & 11.2 GHz \\ \hline
  $|h_{x^\prime}|$ & 1 & 1 \\
  $|h_{y}|$ & 0.02 & 0.14 \\
  $|h_{z^\prime}|$ & 0.19 & 0.37 \\
  $\Phi_{x^\prime}$ & -23$^\circ$ & 50$^\circ$ \\
  $\Phi_{y}$ & 40$^\circ$ & -30$^\circ$\\
  $\Phi_{z^\prime}$ & -33$^\circ$ & 82$^\circ$\\
  \label{table2}
\end{tabular}
\end{table}

The results of the fit have been tabulated in Table \ref{table2}
where $\gamma/2\pi=28.0~ \mu_0$GHz/T, $\mu_0M_0$ = 0.97 T and
$\mu_0H_r$ = 152  mT were used.  The amplitudes of the different
\textbf{h} field components have been normalized with respect to
the $h_{x^\prime}$ component.  At both 11.2 and 12 GHz the
$h_{x^\prime}$ field is much larger than $h_{y}$ or $h_{z^\prime}$, which is expected based on
the wave propagation in a horn antennae.

In changing from 11.2 to 12 GHz the relative phase for each
component is seen to change. Therefore even in the case of a
complex line shape produced by multiple \textbf{h} field
components, by separating the individual contributions of the rf magnetic field via angular dependence measurements, the
relative phase $\Phi$ of each field
component is found to be frequency dependent.

\begin{figure} [h!]
\begin{center}
\epsfig{file=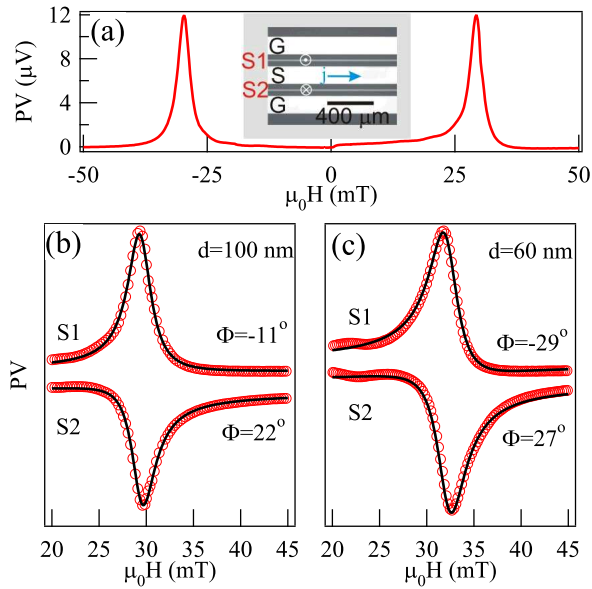,width=8 cm} \caption{(color online). (a) FMR
observed in a first generation spin dynamo.  Inset shows the device structure with two Py strips labeled S1 and
S2.  (b) FMR for Py thickness $d$ = 100 nm for both S1 and S2. In S1
$\Phi= -11^\circ$, while in S2 the line shape is slightly more
asymmetric and $\Phi = 22^\circ$.  (c) For $d$ = 60 nm the relative
phase is $\Phi=-29^\circ$ for S1 and $\Phi=27^\circ$ for S2.}
\label{fig9} \end{center}
\end{figure}

\subsection{Additional Influences on $\Phi$}

In addition to the frequency and sample dependencies, the relative phase $\Phi$ may also depend
on the lead configuration and wiring conditions of a particular
device, as we have
mentioned in Section A. Here, we address such additional influences by using the
first generation spin dynamos\cite{Gui SRE} shown in the inset of Fig.
\ref{fig9}(a). Two spin dynamos with the same lateral dimensions
but different Py thickness $d$ are studied. Each spin dynamo
involves two identical Py strips denoted by S1 and S2, one in each
center of the G-S strips of the CPW, which are placed
symmetrically with respect to the S strip.  The current and rf
$\textbf{h}$ field are induced in the Py via a microwave current
directly injected into the CPW.  Similar to the sample discussed
in Section A, $h_y$ is the dominant field which drives the FMR.

As shown in Fig. \ref{fig9}(a), FMR measured at $\omega/2\pi$ = 5 GHz
on the sample S1 with $d$ = 100 nm shows a nearly symmetric
Lorentz line shape and a field symmetry of $V(H)=V(-H)$. From the
FMR line shape fitting, $\Phi$ = -11$^\circ$ is found.
Interestingly, as shown in Fig. \ref{fig9}(b), the FMR of the sample S2
of the same spin dynamo measured under the same experimental
conditions shows a different line shape, from which a different
$\Phi$ = 22$^\circ$ is found.  We can further compare $\Phi$
measured on the other spin dynamo with a different Py thickness of
$d$ = 60 nm, also at $\omega/2\pi$ = 5 GHz.  Here for S1, $\Phi$ =
-29$^\circ$ while for S2, $\Phi$ = 27$^\circ$.  Again, the relative phase is found different for S1 and S2.
These results demonstrate that due to additional influences such
as a different lead configuration and wiring conditions, even for
samples with the same lateral dimensions, $\Phi$ in each device is
not necessarily the same. It demonstrates clearly that the relative phase $\Phi$ can not be simply determined
by analyzing the FMR line shape measured on a reference device.

\subsection{Closing Remarks}

The experimental data presented above show that regardless of the FMR driving field configuration, the relative
phase between the rf electric and magnetic field is sample and
frequency dependent and non-zero.  This non-zero phase results in
both symmetric and antisymmetric Lorentz line shapes in the FMR detected via field-torque induced spin rectification.
The $\Phi$ dependence of the line shape symmetry changes based on
which component of the rf $\textbf{h}$ field is responsible for
driving the FMR precession.  For instance a purely antisymmetric
line shape could correspond to $\Phi = 0^\circ$ if the
FMR is driven by $h_{x^\prime}$, or to $\Phi=90^\circ$ if the
FMR is driven by $h_{y}$, therefore the line shape itself cannot
be used to determine $\Phi$ directly.  To separate the \textbf{h}
field components an angular ($\theta_H$) dependence measurement is necessary, which
allows both \textbf{h} as well as the phase to be determined.
Using such a measurement $\Phi$ has been observed to change from
0$^\circ$ to 90$^\circ$ in a narrow frequency range (0.6 GHz)
resulting in a change from an antisymmetric to symmetric line
shape demonstrating the large effect the relative phase has on the
FMR line shape.  Furthermore $\Phi$ is not identical even in samples with the same geometrical size.  Therefore
in our opinion $\Phi$ cannot be simply determined from a reference sample but should be calibrated
for each sample, at each frequency and for each measurement cycle.

\section{Conclusion}

Spin rectifications caused by the coupling between current and magnetization in a ferromagnetic
microstrip provide a powerful tool for the study of spin dynamics.
In order to distinguish different mechanisms which enable the electrical detection of FMR via microwave photovoltages, it is essential to properly analyze the FMR line shape. For spin rectification caused by a microwave field torque, due to the coherent
nature of this coupling, the resulting dc voltage depends strongly
on the relative phase between the rf electric and magnetic fields
used to drive the current and magnetization, respectively.
Therefore not only does electrical FMR detection provide a route
to study the relative phase, but it also necessitates calibrating
the relative phase prior to performing electrically detected FMR
experiments. Based on a systematic study of the electrically detected FMR, the
line shape is observed to depend strongly on the microwave
frequency, driving field configuration, sample structure and even wiring conditions. It is in general a combination of
Lorentz and dispersive contributions.  These effects have been quantitatively explained by
accounting for the relative phase shift between electric and
magnetic fields. Analytical formula have been established to analyze
the FMR line shape. Our results imply that for electrically detected FMR which involves both spin Hall and spin rectification effects, the pivotal relative phase must be calibrated
independently in order to properly analyze the FMR line shape and quantify the spin Hall angle. This cannot be done by using a reference sample but could be achieved through such techniques as spintronic Michelson
interferometry\cite{Phase Andre}.

\section*{ACKNOWLEDGEMENTS}
We would like to thank B. W. Southern, A. Hoffmann, and S. D. Bader for discussions. This work has
been funded by NSERC, CFI, CMC and URGP grants (C.-M. H.). ZXC
was supported by the National Natural Science Foundation of China
Grant No. 10990100.


\clearpage

\begin{thebibliography}{199}

\bibitem{Tsoi Nature2000}
M. Tsoi, A. G. M. Jansen, J. Bass, W.-C. Chiang, V. Tsoi, and P.
Wyder, Nature (London) \textbf{406}, 46 (2000).

\bibitem{Ralph Nature2003}
S. I. Kiselev, J. C. Sankey, I. N. Krivorotov, N. C. Emley, R. J.
Schoelkopf, R. A. Buhrman, and D. C. Ralph, Nature (London)
\textbf{425}, 380 (2003).

\bibitem{Tulapurkar Spindiode}
A. A. Tulapurkar, Y. Suzuki, A. Fukushima, H. Kubota, H. Maehara,
K. Tsunekawa, D. D. Djayaprawira, N. Watanabe and S. Yuasa, Nature
(London) \textbf{438}, 339 (2005).

\bibitem{Gui PR}
Y. S. Gui, S. Holland, N. Mecking, and C.-M. Hu, Phys. Rev. Lett.
\textbf{95}, 056807 (2005).

\bibitem{Azevedo spin-pumping}
A. Azevedo, L. H. Vilela Leo, R. L. Rodriguez-Suarez, A. B.
Oliveira, and S. M. Rezende, J. Appl. Phys. \textbf{97}, 10C715
(2005).


\bibitem{VanWees PR}
M. V. Costache, S. M. Watts, M. Sladkov, C. H. van der Wal, and B.
J. van Wees, Appl. Phys. Lett. \textbf{89}, 232115 (2006).

\bibitem{VanWees spin-pumping}
M. V. Costache, M. Sladkov, S. M. Watts, C. H. van der Wal, and B.
J. van Wees, Phys. Rev. Lett. \textbf{97}, 216603 (2006).

\bibitem{Saitoh ISHE}
E. Saitoh, M. Ueda, H. Miyajima, and G. Tatara, Appl. Phys. Lett.
\textbf{88}, 182509 (2006).

\bibitem{NanoFMR}
J. C. Sankey, P. M. Braganca, A. G. F. Garcia, I. N. Krivorotov,
R. A. Buhrman, and D. C. Ralph, Phys. Rev. Lett. \textbf{96},
227601 (2006).


\bibitem{Kubota Spindiode}
H. Kubota, A. Fukushima, K. Yakushiji, T. Nagahama, S. Yuasa, K.
Ando, H. Maehara, Y. Nagamine, K. Tsunekawa, D. D. Djayaprawira,
N. Watanabe, and Y. Suzuki, Nature Physics 4, 37 (2007).

\bibitem{Sankey Spindiode}
J. C. Sankey, Y. T. Cui, J. Z. Sun, J. C. Slonczewski, Robert A.
Buhrman, and D. C. Ralph, Nature Physics 4, 67 (2007).

\bibitem{Gui SRE}
Y. S. Gui,  N. Mecking, X. Zhou, G. Williams and C. -M. Hu, Phys.
Rev. Lett. \textbf{98}, 107602 (2007).


\bibitem{Yamaguchi SRE}
A. Yamaguchi, H. Miyajima, T. Ono, Y. Suzuki, S. Yuasa, A.
Tulapurkar, and Y. Nakatani, Appl. Phys. Lett. \textbf{90}, 182507
(2007).

\bibitem{Goennenwein PR}
S. T. Goennenwein, S. W. Schink, A. Brandlmaier, A. Boger, M.
Opel, R. Gross, R. S. Keizer, T. M. Klapwijk, A. Gupta, H. Huebl,
C. Bihler, and M. S. Brandt, Appl. Phys. Lett. 90, 162507 (2007).

\bibitem{Nikolai2007PRB}
N. Mecking, Y. S. Gui, and C.-M. Hu, Phys. Rev. B \textbf{76},
224430 (2007).

\bibitem{Ralph ST}
J. C. Sankey, Y. T. Cui, J. Z Sun, J. C. Slonczewski, R. A.
Buhrman and D. C. Ralph, Nature Physics, \textbf{4}, 67 (2008)

\bibitem{Xiong Fe-FMR}
X. Hui, A. Wirthmann, Y. S. Gui, Y. Tian, X. F. Jin, Z. H. Chen,
S. C. Shen, and C. -M. Hu,  Appl. Phy. Lett. \textbf{93}, 232502
(2008).

\bibitem{Andre GaMnAs}
A. Wirthmann, X. Hui, N. Mecking, Y. S. Gui, T. Chakraborty, C. -M.
Hu, M. Reinwald, C. Sch\"{u}ller, and W. Wegscheider, Appl. Phys.
Lett. \textbf{92}, 232106 (2008).

\bibitem{Atsarkin LaSrMnO}
V. A. Atsarkin, V. V. Demidov, L. V. Levkin, and A. M. Petrzhik,
Phys. Rev. B \textbf{82}, 144414 (2010).

\bibitem{Mosendz ISHE}
O. Mosendz, J. E. Pearson, F. Y. Fradin, G. E. W. Bauer, S. D.
Bader, and A. Hoffmann, Phys. Rev. Lett. \textbf{104}, 046601
(2010).

\bibitem{Mosendz ISHE2}
O. Mosendz, V. Vlaminck, J. E. Pearson, F. Y. Fradin, G. E. W. Bauer, S. D. Bader, and A. Hoffmann, Phys. Rev. B \textbf{82}, 214403
(2010).

\bibitem{Py/GaAs}
P. Saraiva, A. Nogaret, J. C. Portal, H. E. Beere, and D. A.
Ritchie, Phys. Rev. B \textbf{82}, 224417 (2010).

\bibitem{Saitoh Y3Fe5O12/Pt ISHE}
Y. Kajiwara, K. Harii, S. Takahashi, J. Ohe, K. Uchida, M.
Mizuguchi, H. Umezawa, H. Kawai, K. Ando, K. Takanashi, S.
Maekawa, and E. Saitoh, Nature(London) \textbf{464}, 262 (2010).

\bibitem{Hillebrands Y3Fe5O12/Pt ISHE}
C. W. Sandweg, Y. Kajiwara, K. Ando, E. Saitoh, and B.
Hillebrands, Appl. Phys. Lett. \textbf{97}, 252504 (2010).

\bibitem{Liu SHE}
L. Liu, T. Moriyama, D. C. Ralph, and R. A. Buhrman Phys. Rev.
Lett. \textbf{106}, 036601 (2011).

\bibitem{Azevedo ISHE}
A. Azevedo, L. H. Vilela-Le\~{a}o, R. L. Rodr\'{i}guez-Su\'{a}rez,
A. F. Lacerda Santos, and S. M. Rezende, Phys. Rev. B \textbf{83},
144402 (2011).

\bibitem{Gui Boundary}
Y. S. Gui, N. Mecking, and C.-M. Hu, Phys. Rev. Lett. \textbf{98},
217603 (2007).

\bibitem{Gui Damping1}
Y. S. Gui, A. Wirthmann, N. Mecking, and C.-M. Hu, Phys. Rev. B
\textbf{80}, 060402(R) (2009).

\bibitem{Gui Damping2}
Y. S. Gui, A. Wirthmann, and C.-M. Hu, Phys. Rev. B \textbf{80},
184422 (2009).

\bibitem{ND_Boone}
C. T. Boone, J. A. Katine, J. R. Childress, V. Tiberkevich, A.
Slavin, J. Zhu, X. Cheng, and I. N. Krivorotov, Phys. Rev. Lett.
\textbf{103}, 167601 (2009).

\bibitem{DWR_Bedau}
D. Bedau, M. Kl\"{a}ui, S. Krzyk, U. R\"{u}diger, G. Faini and L.
Vila, Phys. Rev. Lett. \textbf{99}, 146601 (2007).

\bibitem{parametric excitation 1}
S. Bonetti, V. Tiberkevich, G. Consolo, G. Finocchio, P. Muduli,
F. Mancoff, A. Slavin, and J. {\AA}kerman, Phys. Rev. Lett.
\textbf{105}, 217204 (2010).

\bibitem{parametric excitation 2}
S. Urazhdin, V. Tiberkevich, and A. Slavin, Phys. Rev. Lett.
\textbf{105}, 237204 (2010).

\bibitem{RMP2005}
Y. Tserkovnyak, A. Brataas, G. E. W. Bauer, and B. I. Halperin,
Rev. Mod. Phys. \textbf{77}, 1375 (2005).

\bibitem{Wang spin-pumping}
X. Wang, G. E. W. Bauer, B. J. van Wees, A. Brataas, and Y.
Tserkovnyak, Phys. Rev. Lett. \textbf{97}, 216602 (2006).

\bibitem{Demodulation}
A. Yamaguchi, H. Miyajima, S. Kasai, and T. Ono, Appl. Phys. Lett.
\textbf{90}, 212505 (2007).


\bibitem{Phase Andre}
A. Wirthmann, X. Fan, Y. S. Gui, K. Martens, G. Williams, J.
Dietrich, G. E. Bridges, and C.-M. Hu, Phys. Rev. Lett.
\textbf{105}, 017202 (2010).

\bibitem{Phase Zhu}
X. F. Zhu, M. Harder, A. Wirthmann, B. Zhang, W. Lu, Y. S. Gui,
and C.-M. Hu, Phys. Rev. B \textbf{83}, 104407 (2011).

\bibitem{Phase Fan}
X. Fan, S. Kim, X. Kou, J. Kolodzey, H. Zhang, and J. Q. Xiao,
Appl. Phys. Lett. \textbf{97}, 212501 (2010).

\bibitem{Bai2008}
L. H. Bai, Y. S. Gui, A. Wirthmann, E. Recksiedler, N. Mecking,
C.-M. Hu, Z. H. Chen, and S. C. Shen, Appl. Phys. Lett.
\textbf{92}, 032504 (2008).

\bibitem{Zhao Coherent}
H. Zhao, E. J. Loren, H. M. van Driel, and A. L. Smirl, Phys. Rev.
Lett. \textbf{96}, 246601 (2006).

\bibitem{Wang Coherent}
J. Wang, B. F. Zhu, and R. B. Liu, Phys. Rev. Lett. \textbf{104},
256601 (2010).

\bibitem{Werake Coherent}
L. K. Werake, and H. Zhao, Nature Physics, \textbf{6}, 875 (2010).

\bibitem{Jackson}
J. D. Jackson, \textit{Classical Electrodynamics} (John Wiley \&
Sons, New York, 1975), 2nd ed.

\bibitem{Juretschke}
H. J. Juretschke, J. Appl. Phys. \textbf{31}, 1401 (1960).

\bibitem{Silsbee1979}
R. H. Silsbee, A. Janossy, and P. Monod, Phys. Rev. B \textbf{19},
4382 (1979).

\bibitem{Heinrich2003}
B. Heinrich, Y. Tserkovnyak, G. Woltersdorf, A. Brataas, R. Urban,
and G. E. W. Bauer, Phys. Rev. Lett. \textbf{90}, 187601 (2003).

\bibitem{Kupferschmidt2006}
J. N. Kupferschmidt, S. Adam, and P. W. Brouwer, Phys. Rev. B
\textbf{74}, 134416 (2006).

\bibitem{Kovalev2007}
A. A. Kovalev, G. E. W. Bauer, and A. Brataas, Phys. Rev. B
\textbf{75}, 014430 (2007).

\bibitem{Gilbert2004}
T. L. Gilbert, IEEE Trans. Magn. 40, 3443 (2004).

\bibitem{Landau1969}
L. D. Landau and E. M. Lifshitz, \textit{Mechanics}, second edition
(Pergamon Press, Oxford, 1969).

\bibitem{Ohm's-law}
J. P. Jan, in \textit{Solid State Physics}, edited by F. Seitz and
D. Turnbull (Academic, New York, 1957), Vol. 5.

\bibitem{Born1999}
M. Born and E. Wolf, \textit{Principles of optics: Electromagnetic
theory of propagation, interference and diffraction}, 7th edition
(Cambridge university press, Cambridge, 1999).

\bibitem{Heinrich1990}
W. Heinrich, IEEE Trans. Microwave Theory Tech. \textbf{38},
1468 (1990).

\bibitem{Heinrich1993}
W. Heinrich, IEEE Trans. Microwave Theory Tech. \textbf{41},
45 (1993).

\bibitem{FMR1966}
S. V. Vonsovskii, \textit{Ferromagnetic Resonance: The Phenomenon
of Resonant Absorption of a High-Frequency Magnetic Field in
Ferromagnetic Substances}, (Oxford: Pergamon, 1966).

\bibitem{Guru2004}
B. S. Guru, and H. R. Hiziro\v{g}lu, \textit{Electromagnetic Field
Theory Fundamentals}, 2nd ed. (Cambridge University Press,
Cambridge, England, 2004).

\bibitem{Slavin nonlinear damping}
V. Tiberkevich and A. Slavin, Phys. Rev. B \textbf{75}, 014440
(2007).



\end{thebibliography}
\end{document}